\documentclass{pasj00}
\draft

\begin{document}
\SetRunningHead{R. Kitai, H. Watanabe and K. Otsuji}{Penumbral formation}

\title{Morphological study of penumbral formation}

\author{Reizaburo \textsc{kitai} }%
\affil{Kwasan and Hida Observatories, Kyoto University, Yamashina, Kyoto 607-8471, JAPAN}
\email{kitai@kwasan.kyoto-u.ac.jp}

\author{Hiroko \textsc{watanabe}}
\affil{Unit of Synergetic Studies for Space, Kyoto University, Yamashina, Kyoto 607-8471, JAPAN}
\and
\author{Ken'ichi \textsc {otsuji}}
\affil{National Astronomical Observatory of Japan, 2-21-1 Mitaka, Tokyo 181-8588, JAPAN}

%


\maketitle

\begin{abstract}
Penumbrae are known to be area of mainly horizontal magnetic field surrounding umbrae of relatively large and mature sunspots. 
In this paper, we observationally studied the formation of penumbrae in NOAA10978, where several penumbral formations were observed in G-band images of SOT/Hinode. Thanks to the continuous observation by Hinode, we could morphologically follow the evolution of sunspots and found that there are several paths to the penumbral formation: (1) Active accumulation of magnetic flux, (2) Rapid emergence of magnetic field, and (3) Appearance of twisted or rotating magnetic tubes. In all of these cases, magnetic fields are expected to sustain high inclination at the edges of flux tube concentration longer than the characteristic growth time of downward magnetic pumping. 
\end{abstract}

\section{Introduction}
Well-developed sunspots are fledged with surrounding penumbrae. The magnetic field in penumbral area are known to be nearly horizontal (\cite{Stix2002}). The formation mechanism of penumbrae have been studied by many authors theoretically or observationally (\citet{Borrero2011}, \citet{Rempel2011}, and references cited therein).\\   
\indent \citet{Rucklidge1995} studied the energy flows in a model sunspot including the lateral energy flux from surrounding inclined magnetic field, and showed that sunspot MHD structure has two equilibrium states ( pores and sunspots ) and that the structure will change abruptly or transit from pore state to sunspot state when the inclination of surrounding magnetic field gets larger than a threshold value.  Succeeding numerical simulations by \citet{Hurlburt2000} and \citet{Tildesley2004} led to the idea of downward magnetic pumping mechanism around sunspots (\citet{Weiss2004}). \citet{Cheung2010} and then \citet{Rempel2012} closely studied the necessary condition of chromospheric boundary conditions for extended penumbral formation.\\
\indent Observationally, \citet{Shimizu2012} detected a case of penumbral formation, where appearance of chromospheric dark annular zone, which may correspond to horizontal magnetic field, precedes the formation of photospheric penumbra. The work inspired us to observationally follow the evolution of penumbral structure in many varieties of sunspots, continuously in time with Solar Optical Telescope(SOT) on board Hinode (\citet{Suematsu2008}, \citet{Tsuneta2008}). \\  
\indent In this paper, we observationally studied the formation of penumbrae in NOAA10978, where several penumbral formations were observed in G-band images of SOT/Hinode. We will briefly report that there are several paths where the surrounding magnetic field of sunspots are inclined and eventually lead to the formation of penumbrae. 


\section{Morphological study of penumbral formation in NOAA10978 }
\begin{figure}
 \begin{center}
  \includegraphics[width=8cm]{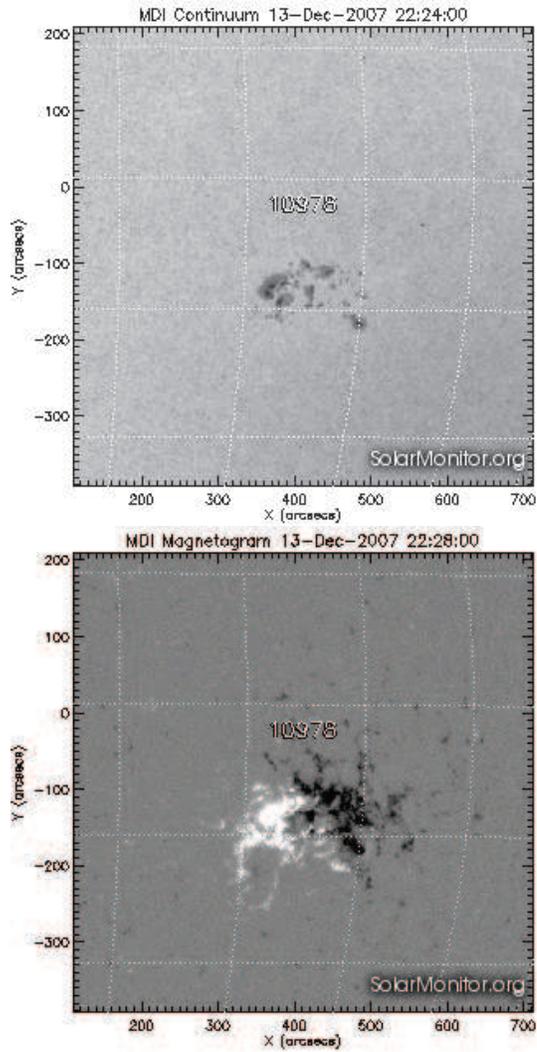} 
 \end{center}
\caption{NOAA10978 on Dec 12, 2007. Top: White light image. Bottom: Map of line-of-sight magnetic field. Courtesy of SolarMonitor.org}\label{fig:SOHO}
\end{figure}

NOAA10978 was a very large sunspot group, covered a wide area on the solar surface, and showed several active emerging flux regions (EFRs) in the region, which developed to form sunspots with penumbrae. Figure \ref{fig:SOHO} show the white light and magnetic maps of the region on Dec 12, 2017.  Flux emergence in the region was very active from Dec 11 through 13. From the visual inspection of the G-band movie, we selected three sub-regions as examples of different paths of penumbral formation. \\

\subsection{Active accumulation of magnetic flux}
\begin{figure}
 \begin{center}
  \includegraphics[width=7cm]{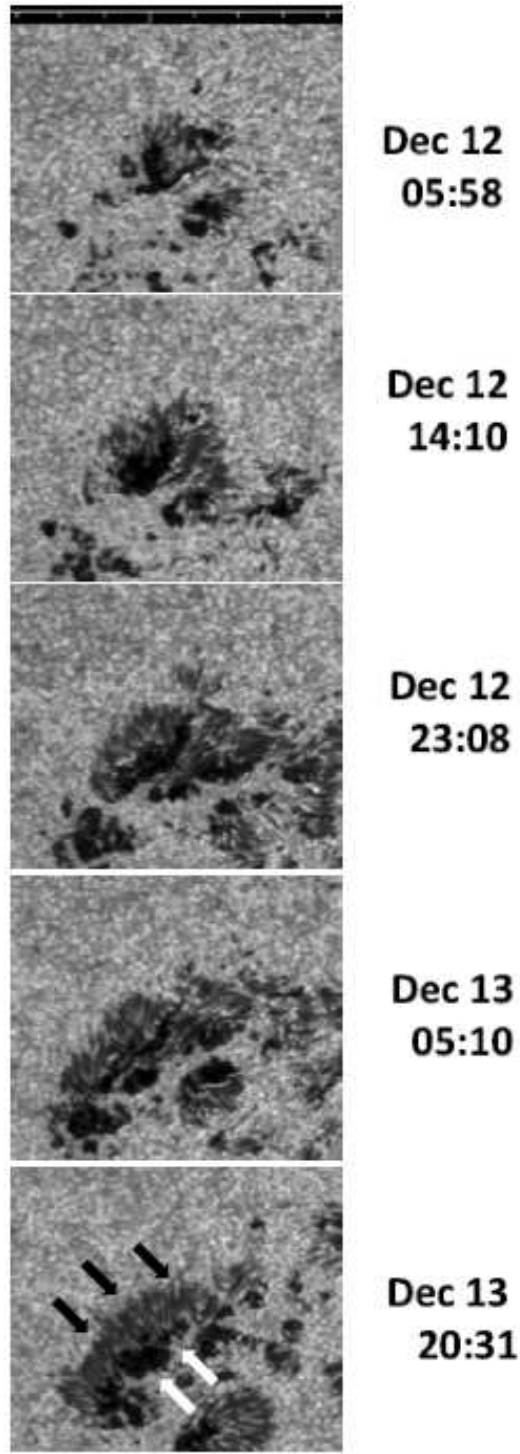} 
 \end{center}
\caption{Rapid accumulation of magnetic flux. The polarity over the area is positive. Black arrows indicate the penumbral area, while white arrows show the umbra photosphere boundary without penumbral structure. The FOV is about 75arcsec$\times$66arcsec.}\label{fig:Rapid_EFR}
\end{figure}

In figure \ref{fig:Rapid_EFR}, we show a case of penumbral formation due to rapid accumulation of magnetic flux. The accumulation continued around 40 hours in this case until the formation of fully developed penumbra. Let us simply assume that the magnetic field of a sunspot takes a canopy structure due to the force balance between the inside magnetic pressure and the outside gas pressure. When the magnetic flux increases, the inclination of magnetic field at the outer periphery of sunspot increases accordingly. Thus, as the magnetic flux is accumulated to the spot, the field will be inclined larger and becomes liable to be horizontal by the downward magnetic pumping.  This case is a typical way of penumbral formation.\\
\indent It is interesting to note that the magnetic polarity is positive all over this particular area.  While northern periphery of the region, indicated by black arrows in figure \ref{fig:Rapid_EFR}, forms penumbral structure, no penumbra are formed at the central part of the umbra, indicated by white arrows.  The latter part is in the midst of positive polarity area, and so the magnetic field will be nearly vertical even at the apparent boundary between the umbra and the normal photosphere, without forming penumbral area.\\

\subsection{Rapid emergence of magnetic field}
\begin{figure}
 \begin{center}
  \includegraphics[width=8cm]{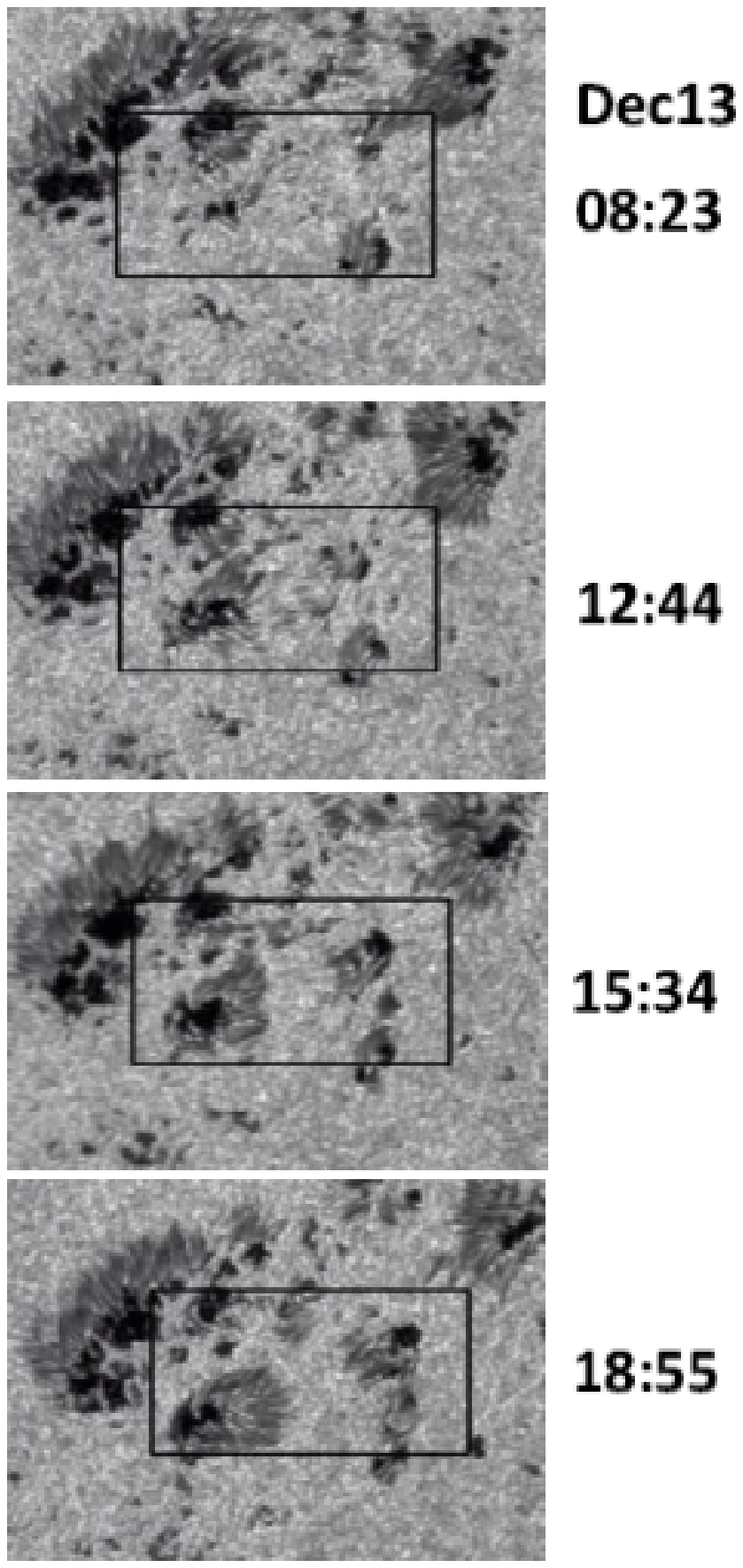} 
 \end{center}
 \caption{Rapidly emerging magnetic flux region. Penumbral area are seen in the middle part of both polarities. The FOV is about 95arcsec$\times$67arcsec.}\label{fig:EFR-2}
\end{figure}

Figure \ref{fig:EFR-2} show the emergence of an EFR at the southern part of the region shown in figure \ref{fig:Rapid_EFR}. The EFR developed very rapidly in 10 hours. In the middle part of the EFR, penumbral structures were seen face to face in both polarities.  Recent observations and numerical simulations show that the emerging magnetic flux stayed horizontally for a while beneath the solar photosphere before the emergence (\citet{Otsuji2011}, \citet{Magara2012}, and \citet{Toriumi2013}). So we can expect that the magnetic field will be nearly horizontal in the middle part of the EFR, which are manifested as photospheric dark lanes often observed in the middle part of EFRs (\citet{Otsuji2007}). Further more, the EFR in concern was emerged at the bottom of pre-existent relatively older bi-polar region and so the expansion of newly emerged magnetic flux to the higher layers will be suppressed.  As the magnetic field is horizontal in the middle part of the EFR, the downward pumping mechanism is highly probable to work in this situation, leading to the formation of penumbral structure.    

\subsection{Appearance of twisted or rotating magnetic tube}
\begin{figure}
 \begin{center}
  \includegraphics[width=7cm]{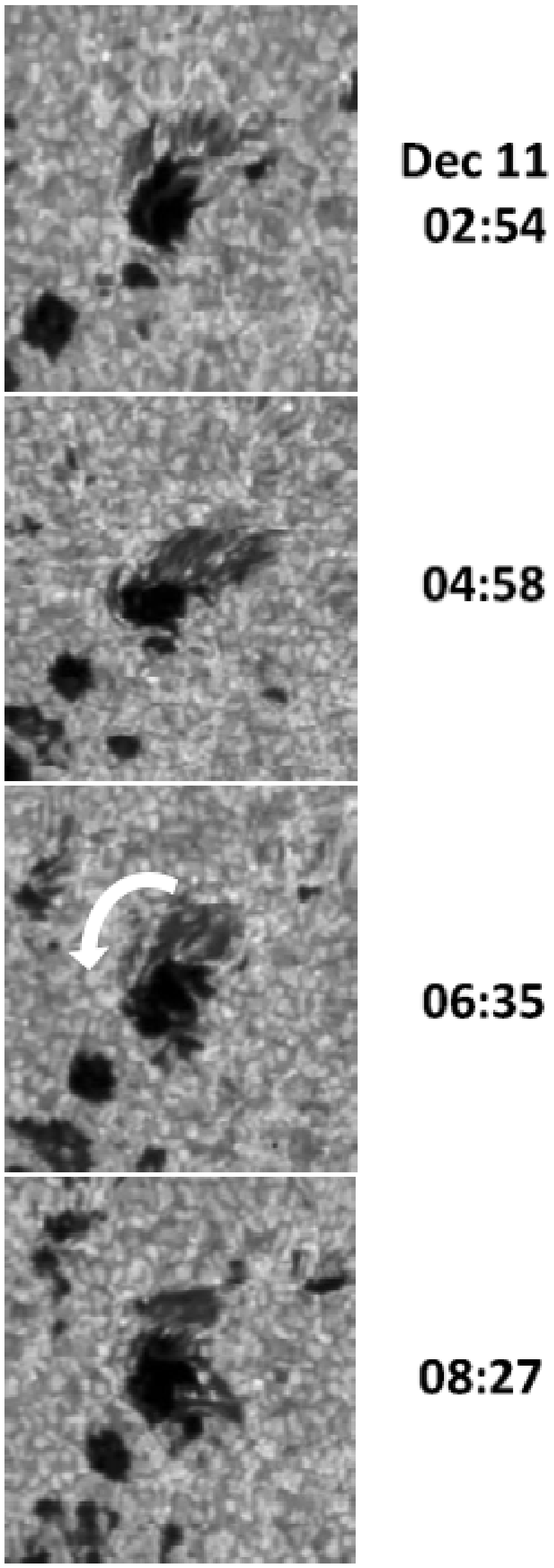} 
 \end{center}
 \caption{Penumbral formation around a rotating or twisted magnetic tube. The umbra rotates counter-clock-wise as is indicated by an arrow. The FOV is about 40arcsec$\times$45arcsec.}\label{fig:Rotation}
\end{figure}

Figure \ref{fig:Rotation} shows a case of penumbral formation where a twisted or rotating magnetic tube appears in the photospheric surface. The apparent counter-clock-wise rotation of the umbra continued 5-6 hours. The situation might be explained by the actually field rotation in counter-clock-wise or by the vertical emergence of twisted field. In either case, the field will be inclined at the umbra-photosphere boundary.  This can be seen from the non-radial pattern of penumbral structure. So in this case, the magnetic field at the periphery of umbra are expected to be highly inclined and is liable to be effected by the magnetic pumping mechanism to form penumbral structure.

\section{Conclusive discussion}
Thanks to the continuous observation by Hinode, we could morphologically follow the evolution of sunspots and found that there are a few different paths to the penumbral formation: (1) Active accumulation of magnetic flux, (2) Rapid emergence of new magnetic flux, and (3) Appearance of twisted or rotating magnetic tubes. In all of these cases, magnetic fields are expected to sustain high inclination at the edges of flux tube concentration longer than the characteristic growth time of downward magnetic pumping ( 3-4 hrs ), which is estimated as 20 times of thermal diffusion time of photospheric convection (\citet{Tildesley2004}). \\
\indent As our report is only based on the morphological study, our findings must be confirmed by the polarimetric observation of three-dimensional magnetic field around penumbra-forming sunspots. Such an observation not only in the formation phases but also in the decaying phases of penumbrae is quite indispensable to clarify the detailed mechanism of penumbral formation.\\
\indent Recently, such studies have been published by several authors. Penumbral formation in the preceeding spot of NOAA11024 was studied by \citet{Schlichenmaier2010a}, \citet{Schlichenmaier2010b} and \citet{Razaei2012} with spectro-polarimetric observation done at the German VTT. The spot grew in area and in the total magnetic flux through the merging of small flux patches. After a few hours of flux merging, penumbral area was formed in the spot at the opposite side of the flux merging. They argue that the total magnetic flux attained a threshold for penumbral formation and the light bridge of inclined magnetic field helped the penumbral formation. As they observed the relatively vertical magnetic field at the merging side and the inclined field at the opposite side, they discussed that these magnetic environment can control the actual penumbral formation.  We think that the penumbral formation in NOAA11024 is a case of active accumulation of magnetic flux and not a case of rapid flux emergence, and agree their discussion that the magnetic environment is a key factor in the actual penumbral formation, as we also notice the one-sided penumbra in Figure \ref{fig:EFR-2}. 
\\
\indent \citet{Louis2013} analyzed the penumbral formation in NOAA11283 with Hinode spectro-polarimetric observation. They found the coalescence of a pore with a decaying sunspot supplied enough magnetic flux and triggered the formation of penumbra at the location of merger, not at the opposite side.  We think that the penumbral formation in NOAA11283 basically is a case of rapid flux accumulation. As the pore was a foot point of an active EFR, we suspect that rapid emergence may play a role in forming the penumbra at the merger side.\\    
\indent \citet{Romano2013} did spectro-polarimetric observation of NOAA11490 and detected the status of the sunspot atmosphere just before the penumbral formation.  They found magnetic canopy structure at the chromospheric heights and uncombed photospheric magnetic structure in the spot atmosphere. We think that the penumbral formation there also is a case of active accumulation of magnetic flux forming a magnetic canopy structure of inclined magnetic field at the periphery. \\
\indent Although these studies contribute much for our understanding of penumbral formation in case of rapid flux accumulation, detailed behavior in the other cases of rapid flux emergence and rotating spots have not been fully observed yet. Farther spectro-polarimetric observation of various cases will be necessary. \\
\color{black}{}

\bigskip 

This work is partly supported by the grant-in-aid from the Japanese MEXT (No 23540264). Hinode is a Japanese mission developed and launched by ISAS/JAXA, with NAOJ as domestic partner and NASA and STFC (UK) as international partners. It is operated by these agencies in co-operation with ESA and NSC (Norway). 



\end{document}